\documentclass{iau} 
\usepackage{graphicx}

\newcommand{\ramses}{\textsc{ramses}}

\title[ICM evolution in stellar clusters]{The evolution of intracluster medium in stellar clusters}

\author[W. Chantereau et al.]{
W. Chantereau,$^{1}$\thanks{E-mail: w.chantereau@ljmu.ac.uk}}

\affiliation{$^{1}$Astrophysics Research Institute, Liverpool John Moores University, 146 Brownlow Hill, Liverpool L3 5RF, UK }

\pubyear{2019}
\volume{351}  
\jname{Star Clusters: From the Milky Way to the Early Universe}
\editors{A. Bragaglia, M.B. Davies, A. Sills \& E. Vesperini, eds.}
\begin{document}

\maketitle
%. CONTINUE EDITING FROM HERE

\begin{abstract} 
Stars in globular clusters lose mass through slow stellar winds that are retained by the stellar cluster and contribute to build up a non negligible intracluster medium over time. However, all the observations so far found only a negligible amount of gas in GCs. We propose here to test different mechanisms such as ram-pressure stripping by the motion of the GC in the galactic halo medium and the inclusion of ionising sources to explain the lack of gas in GCs. We use full 3D hydrodynamical simulations taking into account stellar winds, ionising radiation, radiative heating and radiative pressure. We find that the combining effect of ram-pressure and ionisation are able to explain the negligible amount of gas observed in the core of intermediate-mass and massive GCs.
\end{abstract}


\firstsection % if your document starts with a section,

\section{Introduction}\label{introduction}

Thousands of giant branch stars in globular clusters (GCs) are continuously supplying material through stellar winds that cannot directly escape the host stellar clusters. This material contributes up to $\sim$100-1000~M$_\odot$ to the intracluster medium (ICM) in a few $\sim$100~Myr. However we observe no, or very little, ICM in GCs, for instance \cite[van Loon et al. (2006)]{vanLoon06} observed a 0.3~M$_\odot$ neutral cloud in M15 (NGC~7078) and \cite[Freire et al. (2001)]{Freire01} found that 47~Tuc (NGC~104) contains only $\sim$0.1~M$_\odot$ of gas within its inner 2.5~pc.

Each time GCs cross the galactic disc, it removes all the ICM material from the cluster potential thanks to ram-pressure stripping. However it occurs only every few $\sim$100~Myr. Thus GCs like 47~Tuc which has crossed the Galactic disk around 30~Myr ago \cite[(Gillett et al. 1988)]{Gillett88} should display a non negligible ICM. Therefore several mechanisms have been proposed to remove the gas from GCs between galactic disc crossing to reconcile theory with observations. However, so far, it is has been very difficult to provide a satisfactory mechanism working on short timescales to explain the observations in all GCs with their different properties (cluster mass, distance to the galactic centre, etc.). For instance \cite[Priestley et al. (2011)]{Priestley11} investigated the ram-pressure stripping of the ICM due to the moving GC through the galactic halo and have shown that this process works only for intermediate-mass GCs ($10^5$~M$_\odot$). Recently, \cite[McDonald \& Zijlstra (2015)]{McDonald15} found thanks to hydrostatic model that the photoionisation from WDs and post-AGB stars in stellar clusters would ionise the ICM in every GCs, and in turn, it will expand over the GC's tidal radius and would escape its gravitational potential. 

Here, we follow the suggestion of these recent works of \cite[Priestley et al. (2011)]{Priestley11} and \cite[McDonald \& Zijlstra (2015)]{McDonald15} and investigate by means of full 3D hydrodynamical simulations the combination of the ram-pressure due to the moving GC through the galactic halo and the role played by ionising sources in intermediate-mass and massive GCs ($\sim 10^6$~M$_\odot$). 

\section{Simulation setup}\label{ramses}

We have performed full 3D hydrodynamical simulations using the AMR code \ramses{} \cite[(Teyssier 2002)]{Teyssier02} with and without full radiative transfer \cite[(Rosdahl et al. 2013)]{Rosdahl13} taking into account stellar winds, ionising radiation, radiative heating and radiative pressure.

We reproduce GCs with discretised multi-mass models and their stellar distribution follows a King profile. The stellar clusters evolve in a medium of $10^{5.5}$~K with densities of $6\times10^{-4}$~cm$^{-3}$, $6\times10^{-3}$~cm$^{-3}$ and $7\times10^{-3}$~cm$^{-3}$. The initial metallicity of the gas and stars is $Z = 0.002$. The stellar cluster is in motion through the hot halo medium with an orbital velocity of 200~km.s$^{-1}$ \cite[(Priestley et al. 2011)]{Priestley11}. Each star of the GC on the RGB displays a stellar wind with a velocity of 20~km.s$^{-1}$ \cite[(e.g., McDonald \& van Loon 2007)]{McDonald07} and it leads to a total stellar mass of 3.16 $\times 10^{-7}$~M$_\odot yr^{-1}$ for the intermediate-mass GC and 2.8 $\times 10^{-6}$~M$_\odot yr^{-1}$ for the massive GC. In the simulation where we take into account photoionisation, we include an ionising photon rate of 2.43$\times 10^{47}$ s$^{-1}$ associated to a post-AGB star in the stellar cluster \cite[(McDonald \& Zijlstra 2015)]{McDonald15}.  

We present several simulations to test the effect of different parameters on the ICM evolution. Two simulations follow the gas evolution in intermediate-mass GCs moving through a tenuous ($6\times10^{-4}$~cm$^{-3}$) and dense ($6\times10^{-3}$~cm$^{-3}$) hot galactic halo medium to test the ram-pressure stripping (without photoionisation). Two simulations follow the gas evolution in massive GCs moving in a dense environment with and without an ionising source in the stellar cluster. It allows to test the photoionisation mechanism taking into account full radiative transfer. The properties of the simulations are displayed in Table~\ref{table:simulations}.

\begin{table}
\begin{center}
\begin{tabular}{ c c c c c}
	\hline 
	M$_{cluster}$ (M$_\odot$) & r$_\mathrm{h}$ (pc) & r$_\mathrm{t}$ (pc) & $\dot{M}_{*}$ (M$_\odot yr^{-1}$) & $\rho_{halo}$ (cm$^{-3}$) \\ \hline
    10$^{5}$ & 2.5 & 23.6 & 3.16 $\times 10^{-7}$ & 6 $\times 10^{-4}$-6 $\times 10^{-3}$ \\  
    10$^{6}$ & 7.8 & 52.5 & 2.8 $\times 10^{-6}$ & 7 $\times 10^{-3}$  \\ 
    \hline 
\end{tabular}
\caption[]{Main characteristics of the different simulations. M$_{cluster}$ is the mass of the stellar cluster (M$_\odot$); r$_\mathrm{h}$ the half-mass radius (pc); r$_\mathrm{t}$ the tidal radius (pc); $\dot{M}_{*}$ the total cluster stellar mass-loss rate (M$_\odot yr^{-1}$); $\rho_{halo}$ the halo density (cm$^{-3}$).} \label{table:simulations}
\end{center}
\end{table}

\section{Results}\label{results}

\begin{figure*}
    \centering
    \includegraphics[width=0.70\textwidth]{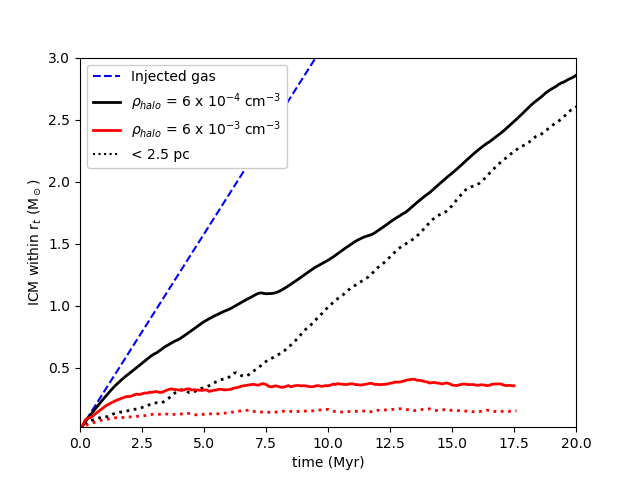} \\  
    \includegraphics[width=0.70\textwidth]{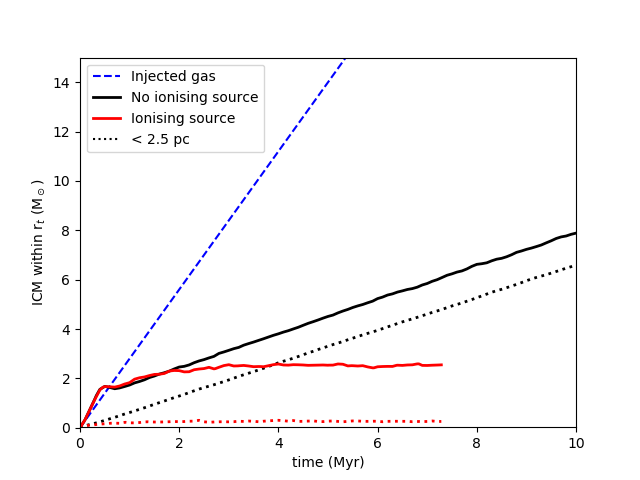} 
    \caption{\textit{Top:} ICM mass within the tidal radius and the central part of 2.5~pc (straight and dotted lines, respectively) for the two simulations of intermediate-mass GCs as a function of time. The simulation in a tenuous halo is in black ($\rho_{halo} = 6\times10^{-4}$~cm$^{-3}$) and the simulation in a denser environment is in red ($\rho_{halo} = 6\times10^{-3}$~cm$^{-3}$). \textit{Bottom:} ICM mass within the tidal radius and the central part of 2.5~pc (straight and dotted lines, respectively) for the two simulations of massive GCs as a function of time. The simulation without an ionising source (no photoionisation) is in black and the simulation with an ionising source (photoionisation) is in red.}
    \label{Figure:mass}
\end{figure*}

\subsection*{Intermediate-mass GC}

We run the first two simulations of intermediate-mass GCs moving in two different halo medium (6$\times 10^{-4}$ and 6$\times 10^{-3}$~cm$^{-3}$) to test the ram-pressure mechanism. We do not take into account any ionising source for these simulations. \\

In a tenuous medium, despite a large part of the gas is stripped from the GC via the ram-pressure mechanism after 20~Myr, most of the ICM is still located in the central region of 2.5~pc of the stellar cluster. Thus even if the ram-pressure stripping allows to get rid of a large part of the ICM, most of it cools and sinks in the inner region of the cluster (Figure~\ref{Figure:mass}). 

In a dense halo medium, most of the cluster is stripped away already at 5~Myr. Only $\sim$0.36~M$_\odot$ is retained within the tidal radius of the cluster and only $\sim$0.15~M$_\odot$ in the central region of the cluster, this value being compatible with observational upper limits found in GCs often lower than 1~M$_\odot$. \\

We conclude that the ram-pressure stripping due to the orbital motion of the GC in a dense halo medium (only) is an efficient mechanism to limit the ICM in the central part of the stellar cluster at levels similar to observations.  

\subsection*{Massive GC}

We run two simulations of massive GCs moving in a dense halo medium with and without an ionising source in the stellar cluster to test the photoionisation mechanism. \\

Without an ionising source, the gas from the stellar winds quickly cools and sinks into the centre of the cluster (6.6~M$_\odot$ within 2.5~pc after only 10~Myr, Figure~\ref{Figure:mass}). Thus even if ram-pressure allows to strip a large part of the gas ($\sim$70~\% of the total injected gas), most of it is still in the central region of the cluster. In massive GCs, the ram-pressure in a dense environment is not sufficient to limit the amount of ICM to levels observed. 

\cite[McDonald \& Zijlstra (2015)]{McDonald15} have shown that UV radiation from hot post-AGB stars and WDs in 47~Tuc can ionise its ICM. Then this ICM will expand until it overflows the GC's tidal radius. Thus here we take into account photoionisation by adding an ionising source in the stellar cluster (hot post-AGB star). The mass accumulated within the tidal radius of the stellar cluster reaches quickly an asymptotic value of 2.55~M$_\odot$. In the central region of the cluster (2.5~pc), the ICM mass is as low as $\sim$0.25~M$_\odot$, similar to the $\sim$0.1~M$_\odot$ of ionised gas determined in the inner part of 47~Tuc \cite[(Freire et al. 2001)]{Freire01}. \\

We can conclude that the photoionisation mechanism is essential in massive GCs and allows with ram-pressure stripping to explain why these stellar clusters have only a negligible amount of ICM. The ionisation mechanism will also be at work in intermediate-mass GCs and then we expect similar results as in massive GCs.

\section{Conclusion}\label{discussion}

We have investigated the ICM evolution in intermediate-mass (10$^5$~M$_\odot$) and massive (10$^6$~M$_\odot$) GCs  with and without ram-pressure stripping and photoionisation. In most cases, the ram-pressure stripping mechanism does not allow to remove all the ICM because the gas cools and sinks into the centre of the GCs. The inclusion of the photoionisation mechanism allows to heat and expand the ICM, in turn, it can be efficiently stripped from the GC by the ram-pressure mechanism.

\end{document}